# Design and Simulation of a Shunt Active Filter in Application for Control of Harmonic Levels

Adrian GLIGOR

Department of Electrical Engineering, Faculty of Engineering,
"Petru Maior" University of Tîrgu Mureş, Tîrgu Mureş, Romania,
e-mail: agligor@upm.ro

**Abstract:** Nowadays, the active filters represent a viable alternative for controlling harmonic levels in industrial consumers' electrical installations. It must be noted the availability of many different types of filter configurations that can be used but there is no standard method for rating the active filters. This paper focuses on describing the shunt active filter structure and design. The theoretical concepts underlying the design of shunt active filters are presented. To validate and highlight the performance of shunt active filters a Matlab-Simulink model was developed. Simulation results are also presented.

## 1. Introduction

After a brief analysis performed on evolution of electric power consumption during the last two decades, it can be observed a change mainly on nature of electric power consumption and profile of consumers. The main causes are represented by introduction of new equipment and facilities to increase comfort in civil construction, new appliances and equipment in order to raise efficiency and diversification of production for industrial consumers, or coexistence in the same building of both households and some industrial consumers. We must also note the impact of the new sources of energy that can easily transform the consumer into power supplier. However, all these changes have led to the emergence of undesirable phenomena in all power system, accounting for the new challenges to be addressed by engineers and scientists involved in the power system design and management.

Among the measures required there must be mentioned the need to adapt the existing electrical network to the new requirements and the introduction of new advanced methods of control, management and monitoring, in order to ensure the efficiency of electricity use.

The aims of this paper are to present a solution to improve the operation of consumers' electrical installations, to reduce the electric power consumption and default costs allocated for the purchase of electricity and removing unwanted effects caused by the presence of harmonics. In order to achieve this, the main goal is to increase the power quality available for consumers. In the case of power consumers affected by the presence of harmonic pollution, power quality improvement can be achieved by implementing systems based on active filtering of the unwanted components. This type of automated system based on shunt active filter is presented in the following sections.

## 2. Operation principle of system based on shunt active filter

*Fig. 1* shows the schematic implementation of active power filter with static power converter.

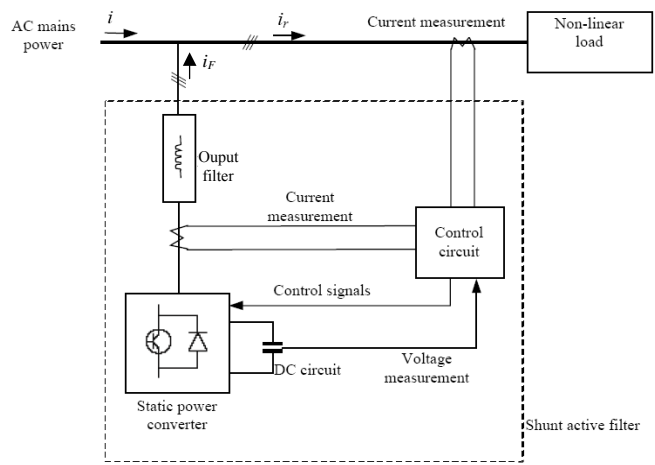

*Figure 1:* The configuration of a system with shunt active filter.

In order to compensate harmonic pollution caused by a nonlinear receiver, the parallel active filter consists of a DC-link static power converter and an energy storage element.

The control circuit performs synthesis of the reference currents of the filter in a manner to compensate the undesired mains current components.

Since currents synthesized by an active filter depend on the average voltage of the storage element, this one should be kept constant. This voltage control has to be provided by the filter control algorithm.

## 3. Structure of the active filter configured for control of harmonics levels

The controller has both the task of controlling the DC-link voltage and the task of controlling the three-phase current system of the active filter.

This requires a complex control structure with two control loops, one for the $i_F$ current, which has to be synthesized and another one for the DC-link voltage.

The structure with two control loops is shown in *Fig. 2*. In this scheme the two controllers can be highlighted: RI – current controller – from the current control loop, and RT voltage controller from the DC-link voltage control loop.

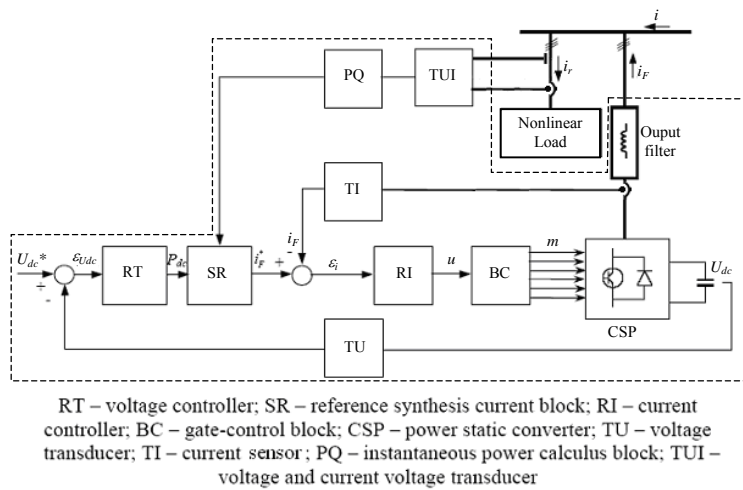

RT – voltage controller; SR – reference synthesis current block; RI – current controller; BC – gate-control block; CSP – power static converter; TU – voltage transducer; TI – current sensor; PQ – instantaneous power calculus block; TUI – voltage and current voltage transducer

*Figure 2:* Block diagram of the system for control of harmonic current level based on active shunt filter.

The voltage controller generates the signal $P_{dc}$ based on the reference signal $U_{dc}*$ and on the feedback signal $U_{dc}$ provided by the voltage sensor TU. SR block, which generates the reference currents, based on signal $P_{dc}$ and on instantaneous powers determined by PQ block, provides the reference for the current control loop. The PQ block has at its input the current and voltage signals provided by the TUI sensor from the circuit which supply the non-linear load.

The RI controller from the current control loop synthesizes the control signal $u$, which is generated from the error signal $\varepsilon_i$. This error signal ($\varepsilon_i$) is obtained by comparing the signal provided by the SR block with the current measured at the input of static power converter (CSP). The control signal $u$ is applied to the BC block, which generates the logic signal $m$ needed to control the CSP block.

*A. Synthesis of references from current compensation control loop based on the instantaneous power theory*

Instantaneous power theory introduced by Akagi offers the methodology for determining the harmonic distortion [1], [2], [3], [4], [5].

According to the notation from *Fig. 1*:

$$
\begin{aligned}
i_{ra} &= I_{r1a}\sqrt{2}\sin\omega t + \tilde{i}_{ra} \\
i_{rb} &= I_{r1b}\sqrt{2}\sin\left(\omega t - \frac{2\pi}{3}\right) + \tilde{i}_{rb}, \\
i_{rc} &= I_{r1c}\sqrt{2}\sin\left(\omega t - \frac{4\pi}{3}\right) + \tilde{i}_{rc}
\end{aligned}
\tag{1}
$$

where $I_{r1a}$ represents the r.m.s value of the load fundamental currents $i_{ra}$, $i_{rb}$, $i_{rc}$, and $\tilde{i}_{ra}$, $\tilde{i}_{rb}$, $\tilde{i}_{rc}$ are the polluting load current components.

In the following there will be noted:

$$
\mathbf{u} = \begin{bmatrix} u_a \\ u_b \\ u_c \end{bmatrix},\ i = \begin{bmatrix} i_a \\ i_b \\ i_c \end{bmatrix},\ \mathbf{i}_r = \begin{bmatrix} i_{ra} \\ i_{rb} \\ i_{rc} \end{bmatrix},\ \mathbf{u}_r = \begin{bmatrix} u_{ra} \\ u_{rb} \\ u_{rc} \end{bmatrix},\ \mathbf{i}_F = \begin{bmatrix} i_{Fa} \\ i_{Fb} \\ i_{Fc} \end{bmatrix}.
\tag{2}
$$

Converting them into *($\alpha$-$\beta$)* coordinates, it results:

$$
\begin{bmatrix} u_{r\alpha} \\ u_{r\beta} \end{bmatrix} = \mathbf{C}\begin{bmatrix} u_{ra} \\ u_{rb} \\ u_{rc} \end{bmatrix},\ \begin{bmatrix} i_{r\alpha} \\ i_{r\beta} \end{bmatrix} = \mathbf{C}\begin{bmatrix} i_{ra} \\ i_{rb} \\ i_{rc} \end{bmatrix},\ \text{where: } \mathbf{C} = \sqrt{\frac{2}{3}}\begin{bmatrix} 1 & -\frac{1}{2} & -\frac{1}{2} \\ 0 & \frac{\sqrt{3}}{2} & -\frac{\sqrt{3}}{2} \\ \frac{1}{\sqrt{2}} & \frac{1}{\sqrt{2}} & \frac{1}{\sqrt{2}} \end{bmatrix}
\tag{3}
$$

Assuming that the zero-sequence components of the three-phase systems are missing, the *C* matrix is given by:

$$\mathbf{C} = \sqrt{\frac{2}{3}} \begin{bmatrix} 1 & -\frac{1}{2} & -\frac{1}{2} \\ 0 & \frac{\sqrt{3}}{2} & -\frac{\sqrt{3}}{2} \end{bmatrix}. \quad (4)$$

In the case of the new two-phase coordinates, the instantaneous power is given by:

$$p = u_{r\alpha} i_{r\alpha} + u_{r\beta} i_{r\beta} \quad (5)$$

If a new quantity is introduced, the so-called instantaneous imaginary power, denoted with $q$, this is defined as (*see Fig. 3*):

$$q\vec{k} = u_{r\alpha}\vec{i} \times i_{r\beta}\vec{j} + u_{r\beta}\vec{j} \times i_{r\alpha}\vec{i} \quad (6)$$

Equation (6) may be rewritten as module as well:

$$q = u_{r\alpha} i_{r\beta} - u_{r\beta} i_{r\alpha}. \quad (7)$$

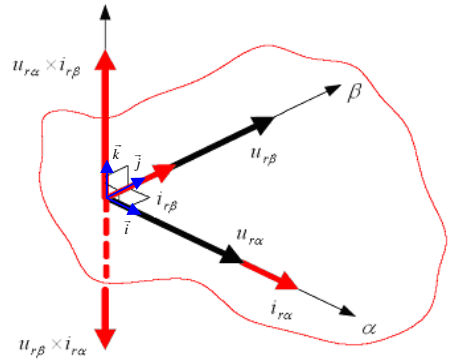

*Figure 3:* Determination of the instantaneous imaginary power.

Equations (5) and (7) may also be rewritten in the form of a matrix as follows:

$$\begin{bmatrix} p \\ q \end{bmatrix} = \begin{bmatrix} u_{r\alpha} & u_{r\beta} \\ -u_{r\beta} & u_{r\alpha} \end{bmatrix} \begin{bmatrix} i_{r\alpha} \\ i_{r\beta} \end{bmatrix}, \quad (8)$$

which results in the expression of current $i_r$ in $(\alpha,\beta)$ system:

$$\begin{bmatrix} i_{r\alpha} \\ i_{r\beta} \end{bmatrix} = \begin{bmatrix} u_{r\alpha} & u_{r\beta} \\ -u_{r\beta} & u_{r\alpha} \end{bmatrix}^{-1} \begin{bmatrix} p \\ q \end{bmatrix} = \frac{1}{u_{r\alpha}^2 + u_{r\beta}^2} \begin{bmatrix} u_{r\alpha} & -u_{r\beta} \\ u_{r\beta} & u_{r\alpha} \end{bmatrix} \begin{bmatrix} p \\ q \end{bmatrix} =$$
$$= \frac{1}{u_{r\alpha}^2 + u_{r\beta}^2} \left\{ \begin{bmatrix} u_{r\alpha} & -u_{r\beta} \\ u_{r\beta} & u_{r\alpha} \end{bmatrix} \begin{bmatrix} p \\ 0 \end{bmatrix} + \begin{bmatrix} u_{r\alpha} & -u_{r\beta} \\ u_{r\beta} & u_{r\alpha} \end{bmatrix} \begin{bmatrix} 0 \\ q \end{bmatrix} \right\} \quad (9)$$

Considering that:

$$p = \bar{p} + \tilde{p} \text{ and } q = \bar{q} + \tilde{q} \quad (10)$$

where:
- $\bar{p}$ represents the component of the instantaneous active power absorbed by the nonlinear load associated to the fundamental of current $i_r$, and voltage $u_r$;
- $\tilde{p}$ represents the component of the instantaneous power absorbed by the nonlinear load associated to the harmonics of current $i_r$ and voltage $u_r$;
- $\bar{q}$ represents the component of the instantaneous imaginary power corresponding to the reactive power associated to the fundamentals of current $i_r$ and voltage $u_r$;
- $\tilde{q}$ represents the component of the instantaneous imaginary power corresponding to the reactive power associated to the harmonics of the current $i_r$ and voltage $u_r$.

If $i_{rA}$ represents the fundamental active component of the absorbed current:

$$\begin{bmatrix} i_{rA\alpha} \\ i_{rA\beta} \end{bmatrix} = \begin{bmatrix} u_{r\alpha} & u_{r\beta} \\ -u_{r\beta} & u_{r\alpha} \end{bmatrix}^{-1} \begin{bmatrix} \bar{p} \\ 0 \end{bmatrix} = \frac{1}{u_{r\alpha}^2 + u_{r\beta}^2} \begin{bmatrix} u_{r\alpha} & -u_{r\beta} \\ u_{r\beta} & u_{r\alpha} \end{bmatrix} \begin{bmatrix} \bar{p} \\ 0 \end{bmatrix} \quad (11)$$

then the reference system of currents in coordinates (α-β) can be obtained in the following form:

$$\begin{bmatrix} i_{F\alpha}^* \\ i_{F\beta}^* \end{bmatrix} = \begin{bmatrix} i_{r\alpha} \\ i_{r\beta} \end{bmatrix} - \begin{bmatrix} i_{rA\alpha} \\ i_{rA\beta} \end{bmatrix} = \begin{bmatrix} u_{r\alpha} & u_{r\beta} \\ -u_{r\beta} & u_{r\alpha} \end{bmatrix}^{-1} \begin{bmatrix} p \\ q \end{bmatrix} - \begin{bmatrix} u_{r\alpha} & u_{r\beta} \\ -u_{r\beta} & u_{r\alpha} \end{bmatrix}^{-1} \begin{bmatrix} \bar{p} \\ 0 \end{bmatrix} =$$
$$= \frac{1}{u_{r\alpha}^2 + u_{r\beta}^2} \begin{bmatrix} u_{r\alpha} & -u_{r\beta} \\ u_{r\beta} & u_{r\alpha} \end{bmatrix} \begin{bmatrix} \tilde{p} \\ \bar{q} + \tilde{q} \end{bmatrix} \quad (12)$$

and the equivalent three-phase components will be, respectively:

$$\begin{bmatrix} i_{Fa}^* \\ i_{Fb}^* \\ i_{Fc}^* \end{bmatrix} = \sqrt{\frac{2}{3}} \begin{bmatrix} 1 & 0 \\ -\frac{1}{2} & \frac{\sqrt{3}}{2} \\ -\frac{1}{2} & -\frac{\sqrt{3}}{2} \end{bmatrix} \begin{bmatrix} i_{F\alpha}^* \\ i_{F\beta}^* \end{bmatrix} \tag{13}$$

*B. Harmonic current compensation by delta modulation*

In case of a hysteresis current control the switching frequency is not well defined. Therefore, it was introduced the concept of the average switching frequency. In principle, the increase of the frequency of power converter leads to a better current compensation. However, the increase of the switching frequency is limited by the switching losses of the power devices.

The operation of the hysteresis current controller is shown in Fig. 4.

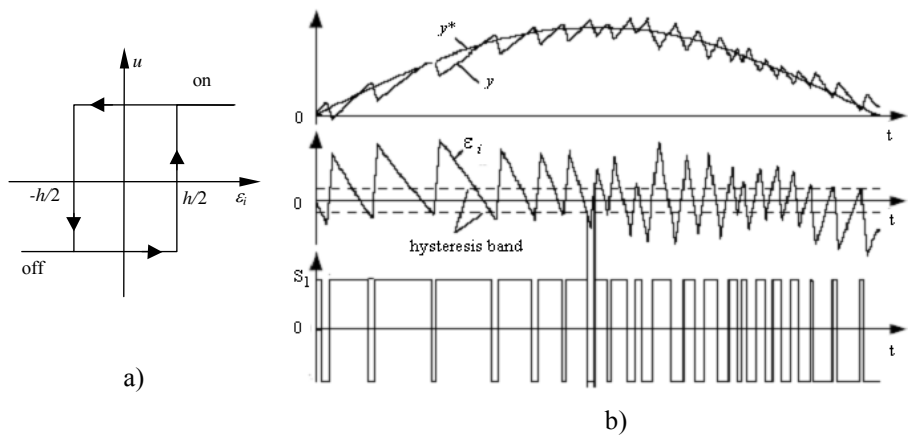

*Figure 4:* Input-output characteristic (a) and the operating principle of the hysteresis controller.

## 4. Results obtained by numerical simulations

The performance analysis of the system with active filtering was realized based on the data obtained by simulation in Matlab-Simulink environment. *Fig. 5* presents the Simulink model used for study.

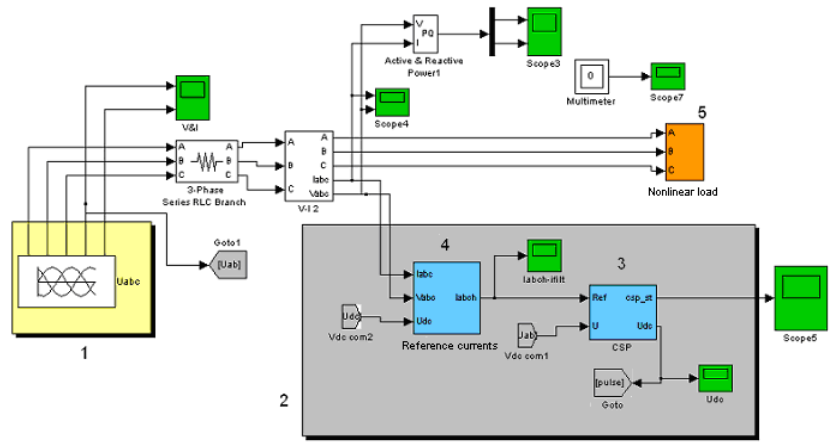

1 – power system, 2 – shunt active filter, 3 – power static converter, 4 – reference current signal generator, 5 – nonlinear load

*Figure 5*: Simulink model.

*Fig. 6* and *8* show the current waveforms and their harmonic spectrum in case of a non-linear load taken for study, whereas *Fig. 7* and *9* present the compensated current waveforms and harmonic spectrum resulted after compensation.

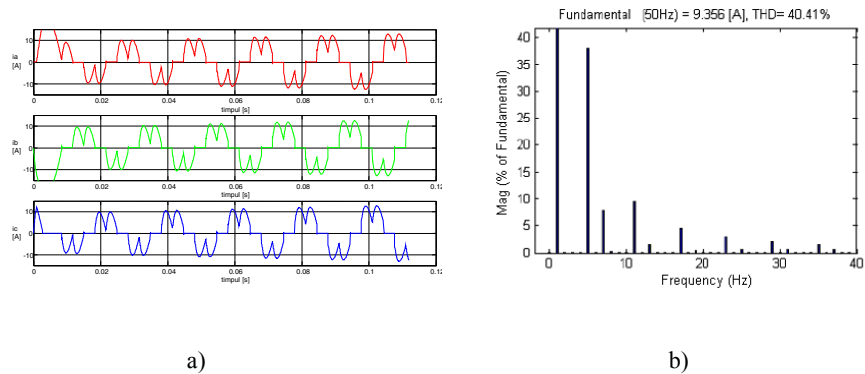

a) b)

*Figure 6:* Current waveforms of a nonlinear load represented by a controlled rectifier in case of a control angle equal with 30°, (a) and the harmonic spectrum of these currents (b).

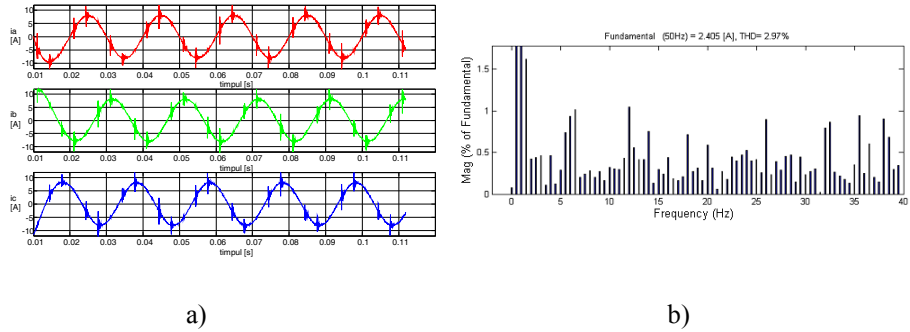

a) b)

*Figure 7*: Compensated mains currents in case of load currents from Fig.6: waveforms (a) and harmonic spectrum (b)

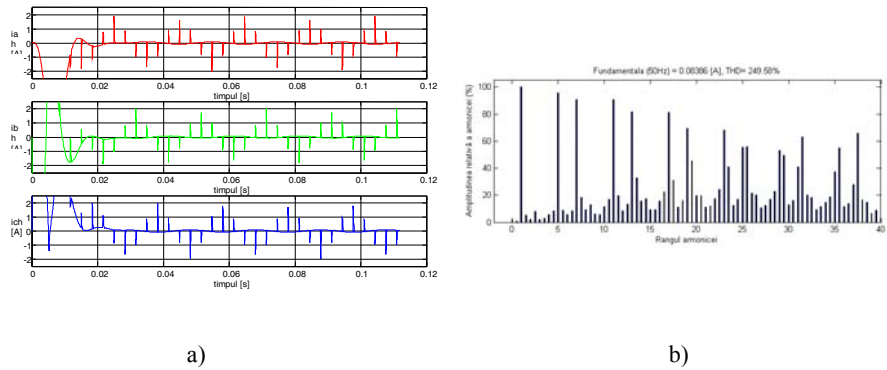

a) b)

*Figure 8:* Current waveforms of a rectifier in case of control angle equal with 10° (a) and the harmonic spectrum of these currents (b).

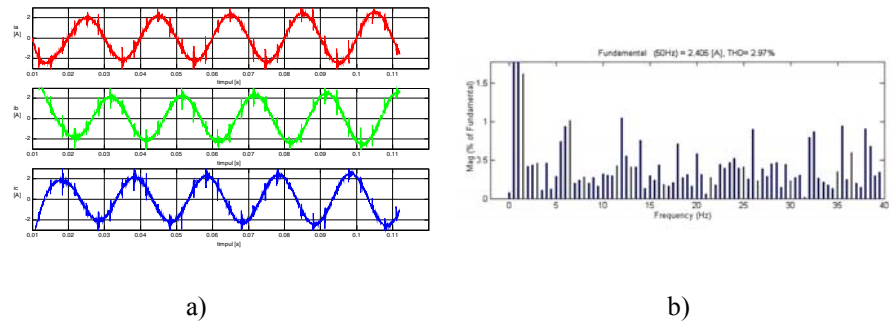

a) b)

*Figure 9*: Compensated mains currents in case of load from Fig. 8: waveforms (a) and harmonic spectrum (b).

In *Table 1, 2* and *3* there are synthesized the main results considering a nonlinear load represented by a controlled rectifier in case of a control angle equal with 30°, and P = 1.54 kW. The automatic system with active filtering achieves a reduction in the level of odd harmonic of 50 ÷97% and a reduction of the $THD_i$ factor by over 94%.

In case of a control angle of 10°, one may observe an increased level of current harmonic distortion ($THD_i$=249.58%). In this case, the automatic system with active filtering achieves a reduction to 2.97% of $THD_i$.

*Table 1: $THD_i$ of the uncompensated system.*

| Phase | $THD_i$ | harmonic ratio [%] | | | | | | | |
|---|---|---|---|---|---|---|---|---|---|
| | | $5^{th}$ | $7^{th}$ | $11^{th}$ | $13^{th}$ | $17^{th}$ | $19^{th}$ | $23^{th}$ | $25^{th}$ |
| A | 53,47 | 48,55 | 18,38 | 9,82 | 5,91 | 3,94 | 2,69 | 2,06 | 1,44 |
| B | 53,55 | 48,58 | 18,56 | 9,79 | 5,96 | 3,90 | 2,74 | 2,04 | 1,48 |
| C | 53,44 | 48,55 | 18,31 | 9,86 | 5,85 | 3,93 | 2,67 | 2,06 | 1,41 |

*Table 2: $THD_i$ of the compensated system.*

| Phase | $THD_i$ | harmonic ratio [%] | | | | | | | |
|---|---|---|---|---|---|---|---|---|---|
| | | $5^{th}$ | $7^{th}$ | $11^{th}$ | $13^{th}$ | $17^{th}$ | $19^{th}$ | $23^{th}$ | $25^{th}$ |
| A | 2,74 | 1,55 | 1,33 | 0,18 | 0,17 | 0,27 | 0,11 | 0,16 | 0,25 |
| B | 2,82 | 1,34 | 1,33 | 0,13 | 0,41 | 0,11 | 0,33 | 0,20 | 0,11 |
| C | 2,50 | 1,08 | 1,43 | 0,29 | 0,56 | 0,21 | 0,44 | 0,17 | 0,35 |

*Table 3:* Reduction of $THD_i$ in case of using compensation [%].

| Phase | Reduction of $THD_i$ | reduction of the harmonic components [%] | | | | | | | |
|---|---|---|---|---|---|---|---|---|---|
| | | $5^{th}$ | $7^{th}$ | $11^{th}$ | $13^{th}$ | $17^{th}$ | $19^{th}$ | $23^{th}$ | $25^{th}$ |
| A | 94,88 | 96,81 | 92,76 | 98,17 | 97,12 | 93,15 | 95,91 | 92,23 | 82,64 |
| B | 94,73 | 97,24 | 92,83 | 98,67 | 93,12 | 97,18 | 87,96 | 90,20 | 92,57 |
| C | 95,32 | 97,78 | 92,19 | 97,06 | 90,43 | 94,66 | 83,52 | 91,75 | 75,18 |

## 5. Conclusion

Proliferation of the power electronic equipments leads to an increasing harmonic contamination in power transmission or distribution systems. Many researchers from the field of the power systems and automation have searched for different approaches to solve the problem. One way was open by introducing the harmonic compensation by using active filters.

This paper presents an automatic system based on active filtering for harmonic current reduction with direct applicability in the civil and industrial electrical installations affected by harmonics.

The proposed system is based on the new theory of instantaneous power introduced by Akagi and on delta modulation control technique of the static power converter.

Simulation results were obtained before and after the use of the automatic system based on active filtering. From the analysis of the experimental data, in case of a nonlinear load of rectifier type, one may observe that there are different levels of current distortion produced depending on the load and its control mode, with high values of the total current harmonic distortion and low power factor.

Using the active filter, the experimental data show that the total harmonic distortion of current ($THD_i$) decreases to 1%-4%, and the power factor rises up to 0.98-1. A 30% decrease of the r.m.s. value of the current was also recorded.

Analyzing the harmonic spectra of the compensated currents, it results that the weight of the 5th and 7th harmonics are close to 1% and the rest of upper harmonics are below 1%.

**References**

[1] Akagi, H., "Modern active filters and traditional passive filters"; *Bulletin of The Polish Academy of Sciences Technical Sciences*; Vol. 54, No. 3, pp. 255-269; 2006.

[2] Akagi, H., "New trends in active filters for improving power quality", in *Proc. of the International Conference on Power Electronics, Drives and Energy Systems for Industrial Growth, 1996*,Vol. 1, Issue 8-11, Jan. 1996, pp. 417 - 425.

[3] Gligor, A., "Contribuții privind sistemele avansate de conducere și optimizare a proceselor energetice în instalațiile electrice la consumatori", *PhD Thesis,* Universitatea Tehnică Cluj-Napoca, 2007.

[4] Codoiu, R., "Selection of the representative non-active power theories for power conditioning", *Proceedings of the International Scientific Conference Inter-Ing. 2007*, Tg. Mureș, 2007, pp.V-6-1 - V-6-14.

[5] Codoiu, R., Gligor, A., "Current and power components simulation using the most recent power theories", *Proceedings of the International Scientific Conference Inter-Ing. 2007*, Tg. Mureș, 2007, pp.V-7-1 - V-7-8.